\documentclass{optica-article}

\journal{opticajournal} 

\articletype{Research Article}

\usepackage{lineno}
\usepackage{float}
\usepackage{upgreek}
\usepackage{multirow}

\begin{document}

\title{Responsivity evaluation of photonics integrated photodetectors via pairwise measurements with an attenuation circuit}

\author{Jing Zhang\authormark{1,*}, Tianchen Sun\authormark{2}, Mai Ji\authormark{2}, Anirudh. R. Ramaseshan\authormark{3,5}, Aswin. A. Eapen\authormark{3}, Thomas Y. L. Ang\authormark{4}, and Victor Leong\authormark{3,5,$\dagger$}}

\address{\authormark{1}National Metrology Centre (NMC), Agency for Science, Technology and Research (A$^\ast$STAR), 8 Cleantech Loop, \#01-20, Singapore 637145, Republic of Singapore\\
\authormark{2}School of Electrical and Electronic Engineering, Nanyang Technological University, 50 Nanyang Ave, Singapore 639798, Republic of Singapore\\
\authormark{3}Institute of Materials Research and Engineering (IMRE), Agency for Science, Technology and Research (A$^\ast$STAR), 2 Fusionopolis Way, Innovis \#08-03, Singapore 138634, Republic of Singapore\\
\authormark{4}Institute of High Performance Computing (IHPC), Agency for Science, Technology and Research (A$^\ast$STAR), 1 Fusionopolis Way, \#16-16 Connexis North Tower, Singapore 138632, Republic of Singapore\\
\authormark{5}Quantum Innovation Centre (Q.InC), Agency for Science, Technology and Research (A$^\ast$STAR), 2 Fusionopolis Way, Innovis \#08-03, Singapore 138634, Republic of Singapore}

\email{\authormark{*}zhang\_jing@nmc.a-star.edu.sg $\dagger$victor\_leong@imre.a-star.edu.sg} 


\begin{abstract*}
Integrated photonics platforms offer a compact and scalable solution for developing next-generation optical technologies. 
For precision applications involving weak signals,
the responsivity as well as the accurate calibration of the integrated photodetectors 
at low optical powers become increasingly important.
It remains challenging to perform a calibration traceable to mW-level primary standards
without relying on external attenuation setups.
Here, we utilize an on-chip attenuation circuit, composed of a series of cascaded directional couplers (DCs), 
to evaluate the responsivity of integrated photodetectors (PDs) at $\sim$$\upmu$W optical power levels with $\sim$mW inputs to the chip. 
Moreover, we show that a pairwise measurement method, 
involving the simultaneous measurement of the integrated PD photocurrent 
and an auxiliary optical output which is coupled off-chip,
systematically improves the experimental uncertainties 
compared to a direct PD photocurrent measurement.
For 3 cascaded DCs, 
the pairwise measurement improves the repeatability error from 1.21\% to 0.22\%,
with an overall expanded calibration uncertainty (k\,=\,2) of 10.13\%.
The latter is dominated by the scattering noise floor and fiber-to-chip coupling errors, which 
can be significantly improved with better device fabrication control.
Our method can be extended to a fully integrated calibration solution for waveguide-integrated single-photon detectors.

\end{abstract*}

\section{Introduction}
\label{sec:Introduction}
Integrated photonics platforms hold significant potential for developing in cutting-edge technologies, 
such as quantum sensing~\cite{pirandola2018advances,chang2023nanowire,cesar2024towards}, 
quantum information~\cite{flamini2018photonic,luo2023recent,labonte2024integrated},
and LIDAR~\cite{9875979,lukashchuk2024photonic}. 
A significant advantage of such an approach is the ability of a fully-equipped photonics platform 
to combine multiple device functionalities in a single device, 
e.g. photon sources, detectors, and fast modulators.
Their compatibility with semiconductor manufacturing processes also provides a scalable pathway towards cost-effective fabrication~\cite{stojanovic2018monolithic}. 

Integrated photodetectors are a critical component of any photonics platform;
devices based on material platforms such as Si~\cite{zhu2017all,chatterjee2019high}, Ge~\cite{liu2022high,wang202064,srinivasan202027}, and III-V~\cite{xue2022high,tossoun2019indium}
have been developed, and some are already being offered in commercial integrated photonics foundries.
However, increasingly precise applications would require higher detection sensitivities, even down to single-photon levels.
There are several major types of single-photon detectors compatible with integrated photonics that are being developed,
including superconducting nanowire single-photon detectors (SNSPDs)~\cite{reddy2020superconducting, colangelo2023impedance, esmaeil2021superconducting} 
and single-photon avalanche detectors (SPADs)~\cite{yanikgonul2021integrated,yanikgonul2022high, martinez2016high}. 
Compared to superconducting photodetectors, SPADs have more modest light detection performance, 
but they do not require cryogenic temperatures to operate and are thus more practical to deploy in certain applications~\cite{liang2016room, ceccarelli2021recent, na2024room}. 

For applications requiring high measurement accuracy,
the integrated photodetectors should undergo proper calibration prior to their deployment.
However, as the calibration is typically performed against a $\sim$mW light source 
that is traceable to a primary standard such as an optical cryogenic radiometer,
the calibration at low optical power levels presents several challenges.
First, accurately obtaining a low-power input source from the primary standard typically requires a bulky off-chip attenuation setup, which can hinder the scalability of integrated photonics devices \cite{karabchevsky2020chip,shaker2023integrated,najafi2015chip}.
Moreover, the low signal-to-noise ratio for weak optical signals increases the measurement uncertainties.

We have previously demonstrated a photonics integrated solution for performing the optical power attenuation on-chip,
using an attenuation circuit based on a series of cascaded directional coupler (DC) structures~\cite{zhang2024integrated}.
Each DC stage provides an attenuation based on its cross coupling transmission factor;
by engaging varying numbers of DC stages, a wide range of optical powers can be obtained from a single optical input power.

Here, we utilize such a photonics attenuation circuit to systematically evaluate the responsivity of an integrated silicon PIN photodetector (PD)
at optical powers down to $\sim$$\upmu$W levels from a $\sim$mW input at 685\,nm wavelength.
We also demonstrate a pairwise measurement method for evaluating the integrated PD, 
where we simultaneously measure the PD photocurrent as well as 
an optical output coupled off-chip via an auxiliary output of the DC.
We show that the pairwise method yields lower measurement uncertainties compared to a direct photocurrent measurement,
as the errors in input power fiber-to-chip coupling are mitigated.
Our method can be extended for the characterization and calibration on photonics-integrated single-photon detectors.

\section{Methods}
\subsection{Photonic circuit design}
\label{sec:Photonic circuit design}
We implement our cascaded DC attenuation circuit with a silicon nitride (SiN) photonics circuit, 
which is integrated with a Si PD based on a PIN-doped silicon rib waveguide on a silicon-on-insulator (SOI) platform.
In the visible and near-infrared wavelength ranges, SiN waveguides exhibit excellent transmittivity with low propagation loss, 
while Si PDs are highly efficient at detecting visible light.
The PD is connected to the SiN photonics circuit via end-fire coupling on the same device layer.
The details of the PD design and overall device fabrication is similar to our previous reports~\cite{yanikgonul2021integrated, gundlapalli2022visible}.

The SiN photonics circuit consists of rectangular waveguides with a width of 400\,nm and height of 340\,nm,
which are single-mode at the target wavelength of 685\,nm.
The bottom and top SiO$_2$ cladding thickness is 2\,$\upmu$m and 3\,$\upmu$m, respectively.
Inverse-tapered edge couplers with a minimum width of 250\,nm allow for efficient light coupling to lensed optical fibers.

\begin{figure}[H]
\centering
\includegraphics[width=0.9\linewidth]{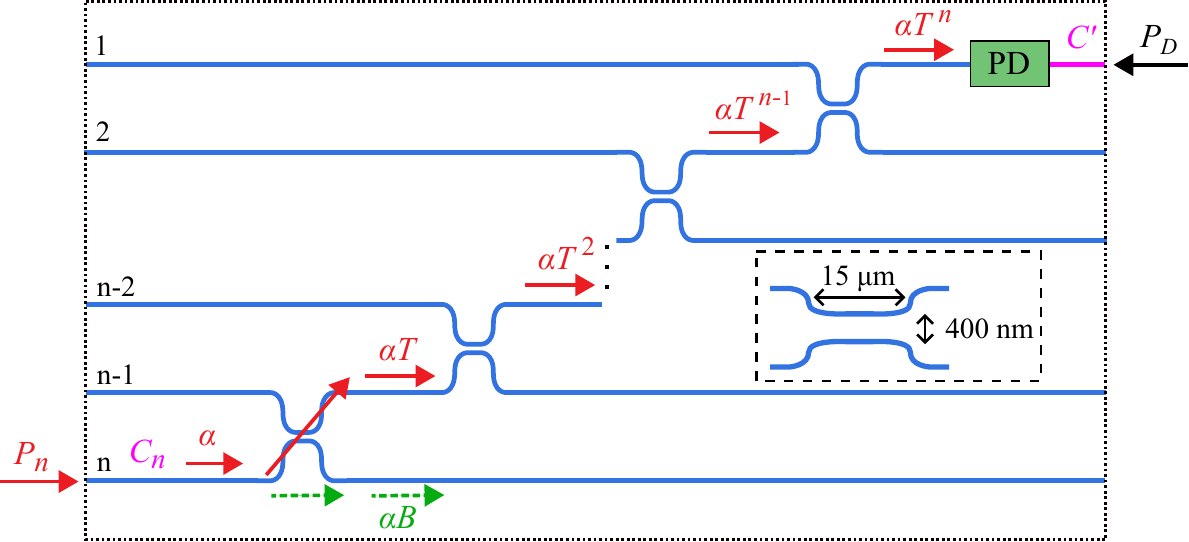}
\caption{Device schematic of the integrated attenuation circuit based on cascaded directional couplers (DCs). 
At the input waveguide $n$, an on-chip input optical power of $\alpha=P_nC_n$ is attenuated to $\alpha T^n$ (red arrows) 
at the photodetector (PD) by $n$ DC stages each having a cross coupling transmission factor of $T$. 
The DC bar coupling factor is~$B$.
To measure the PD directly without using the DCs,
an optical input can also be introduced from the opposite end of the PD (black arrow). 
$C_n$ and $C'$ are the respective coupling efficiencies that include fiber-to-chip coupling and waveguide propagation loss.
The inset depicts the DC coupling length of 15\,$\upmu$m and gap of 400\,nm.  
}
\label{fig:DC_cross_bar}
\end{figure}

Each DC consists of a pair SiN waveguides brought close together with a coupling length of 15\,$\upmu$m and a gap of 400\,nm. 
For an on-chip optical power $\alpha$ incident on the DC, the output at the cross and bar output ports are given by $\alpha T$ and $\alpha B$, respectively
(see Fig.~\ref{fig:DC_cross_bar}), where $T$ is the DC cross coupling transmission factor, $B$ is the bar coupling factor,
and $T+B+\varepsilon=1$ with $\varepsilon$ accounting for the DC insertion loss.
By cascading multiple DC stages via their cross couplings, we can vary the overall attenuation over a large range to obtain the desired optical power level at the PD (see Fig.~\ref{fig:DC_cross_bar}). 

We highlight here the advantage of each DC being a symmetric 2×2 port device: the auxiliary input and output ports can be used to independently characterize the DCs without interfering with the rest of the photonic circuit, 
and will also play an important role in the pairwise measurement method, which is elaborated in the later sections.

\subsection{Responsivity evaluation}\label{sec:Responsivity evaluation}

The responsivity of a photodetector (PD) is $R=I/P_\textrm{in}$,
where $I$ is the photocurrent $I$ and $P_\textrm{in}$ is the optical power incident onto the PD.
We will compare two methods for measuring the responsivity $R$.

\subsubsection{Direct photocurrent measurement}\label{sec:direct}
The most direct way to characterize an integrated PD is to measure the photocurrent $I$ when light is coupled from an off-chip source to the PD via an input waveguide, 
without going through other on-chip photonics structures (Fig.~\ref{fig:DC_cross_bar}).
We perform the optical power attenuation and control off-chip with an external setup.
The PD responsivity is

\begin{equation}
    R=\frac{I}{P_D\times C'}
    \label{eq:for single input responsivity}
\end{equation}
where $P_D$ is the optical power from the input optical fiber, 
and $C'$ is a coupling efficiency term consisting of 
fiber-to-chip coupling and the propagation loss over the relevant waveguide segment (length\,=\,0.7\,mm). 

The major uncertainty sources for method are
the accuracy of the input power $P_D$ and the fiber-to-chip coupling accuracy.
Both are challenging at highly attenuated optical power levels, leading to increased uncertainties.

\subsubsection{Pairwise measurement method}

\begin{figure}[H]
\centering
\includegraphics[width=0.9\linewidth]{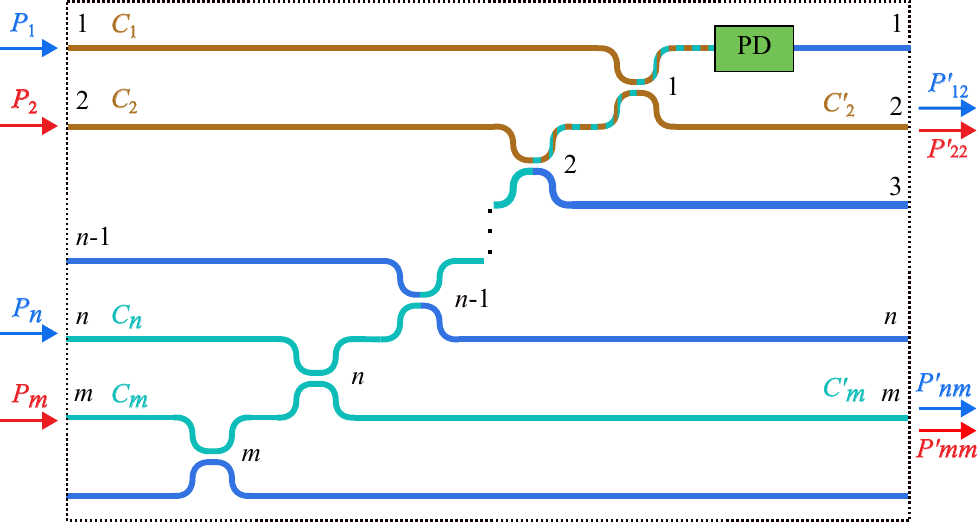}
\caption{Schematic of the pairwise method for measuring the integrated photodetector (PD) responsivity. 
Input light $P$ is sent through a pair of adjacent waveguides $n$ and $m=n+1$.
The light incident on the PD is attenuated by $n$ DC stages. 
The PD photocurrent is measured simultaneously with the optical output $P'$ of waveguide $m$.
The subscripts of $P$ and $P'$ label the input and output waveguides,
and the color coding highlights the optical paths.
$C$ and $C'$ are the respective coupling efficiencies. 
}
\label{fig:for pair input responsivity}
\end{figure}

The pairwise measurement leverages on the dual output ports of the DC
to enable the simultaneous measurement of the PD photocurrent
as well the the auxiliary optical output which is coupled off-chip.
The input and output waveguides, as well as the DCs, are labeled sequentially starting from the position of the PD (see Fig.~\ref{fig:for pair input responsivity}).
The input optical powers $P_n$ and $P_m$ are launched from the fiber into two adjacent input waveguides $n$ and $m$ consecutively, where $m=n+1$,
and the respective auxiliary output powers $P'_{nm}$ and $P'_{mm}$ and are measured at the output waveguide $m$:
\begin{align}
   P'_{nm} &=P_n \times C_n \times T_n \times C'_m\label{eq:for output power Pnm}\\
   P'_{mm} &=P_m \times C_m \times B_m \times B_n \times C'_m\label{eq:for output power Pmm}   
\end{align}
Here, $T_n$ and $B_n$ denote the cross coupling transmission and bar coupling factor of the $n$-th DC, respectively.
The coupling efficiency terms $C$ and $C'$ refer to the input and output sides, respectively,
consisting of the fiber-to-chip coupling as well as the propagation loss of the relevant waveguide segment from the chip edge to the DC.
We note that the $B_m$ term does not apply if the $n$-th DC is the last DC in the cascade (i.e. there is no $m$-th DC). 

Similarly, the PD photocurrent can be written as:
\begin{align}
    I_n &=P_n \times C_n \times B_n \times \prod_{i=1}^{n-1} T_i \times R\label{eq:for photocurrent In}\\
    I_m &=P_m \times C_m \times B_m \times \prod_{i=1}^n T_i \times R\label{eq:for photocurrent Im}
\end{align}
where $I_n$ and $I_m$ are the PD photocurrent when the input light is sent to the corresponding input waveguide. By combining Eq.~(\ref{eq:for output power Pnm})-(\ref{eq:for photocurrent Im}), the responsivity of PD can be derived as:

\begin{equation}
    R = \sqrt{\frac{I_m \times I_n}{P'_{mm} \times P'_{nm}} } \times \frac{C'_m}{\prod_{i=1}^{n-1} T_i}\label{eq:for pair input reponsivity with cascaded DCs}
\end{equation}
For the first pair of waveguides (i.e. $n=1$), this becomes:
\begin{equation}
    R=\sqrt{\frac{I_1 \times I_2}{P'_{12} \times P'_{22}}} \times C'_2\label{eq:for pair input responsivity}
\end{equation}

If all the cascaded DCs have the same design parameters,
and good device fabrication control yields uniform values of $T$ across the DCs within a tight tolerance,
it is possible to simplify Eq.~(\ref{eq:for pair input reponsivity with cascaded DCs}) to
\begin{equation}
    R=\sqrt{\frac{I_m\times I_n}{P'_{mm}\times P'_{nm}}} \times \frac{C'_m}{T^{n-1}}
    \label{eq:for simplified pair input responsivity with cascaded DCs}
\end{equation}

\subsection{DC evaluation}
\label{sec:DC evaluation}

We can also apply the pairwise measurement concept to characterize the DCs, 
by sending input light $P$ into each input waveguide consecutively, 
and measuring the optical outputs $P'$ at both output ports.
For the $n$-th DC, the four measurements can be described by 
(refer to Fig.~\ref{fig:for pair input responsivity} for waveguide labeling): 
\begin{align}
    P'_{nn} &= P_n \times C_n \times B_n \times B_{n-1} \times C'_n\label{eq:for output Pnn} \\
    P'_{nm} &= P_n \times C_n \times T_n \times C'_m\label{eq:for output Pnm} \\
    P'_{mn} &= P_m \times C_m \times B_m \times T_n \times B_{n-1} \times C'_n\label{eq:for output Pmn} \\
    P'_{mm} &= P_m \times C_m \times B_m \times B_n \times C'_m\label{eq:for output Pmm}
\end{align}
We note again that the $B_m$ terms apply only if the $m$-th DC exists. Rearranging, we obtain the simplified expression:
\begin{equation}
    \left(\frac{T_n}{B_n}\right)^2 = \frac{P'_{mn} \times P'_{nm}}{P'_{nn} \times P'_{mm}}
    \label{eq:for DC evaluation}
\end{equation}
This ratio of the cross and bar coupling terms, or cross-bar ratio,
is a quantity that describes the performance of the DC, requiring only the four measured optical power values,
with the coupling efficiency terms (and their associated uncertainties) being canceled out.
As such, the distribution of measured cross-bar ratio values can be used as a convenient and accurate way to evaluate the performance of the DCs,
and hence infer the device fabrication uniformity across a device or wafer.

If the DC insertion loss $\varepsilon$ is negligible, 
the cross-bar ratio can be written as 
$(T_n/(1-T_n))$,
allowing us to extract the cross coupling transmission $T_n$ directly from the measurement of the DC itself via Eq.~\ref{eq:for DC evaluation}.

\subsection{Propagation loss and fiber-to-chip coupling loss}

The waveguide propagation loss and fiber-to-chip coupling losses are required to evaluate the coupling efficiency terms $C$ and $C'$.
We obtain these quantities with the cutback method by measuring the transmission through a set of reference waveguides of different lengths,
and performing a linear regression analysis.
The design parameters of the waveguides and edge couplers are identical for the attenuation circuit and the reference waveguides.

\section{Measurement results}

A 685 nm laser (Thorlabs) is coupled into a polarization-maintaining (PM) input lensed fiber (OZ Optics) and launched into the on-chip waveguide with TE polarization.
The optical output is coupled into a single-mode lensed fiber (OZ Optics) and measured with an optical power meter (Thorlabs).
The fiber-to-chip alignment is optimized with piezoelectric stages running an automated algorithm.
Electrical connections to the PD are established via DC probes (Formfactor) and the on-chip contact pads.
The PD is operated under zero bias voltage, and the current is measured with a sourcemeter (Keithley).

The measured fiber-to-chip coupling loss is -6.2\,$\pm$\,0.2\,dB,
and the waveguide propagation loss is 0.30\,$\pm$\,0.02\,dB/mm
across the wafer.
We thus obtain $C'=10^{(-6.2-0.3\times 0.7)/10}$ = 0.228,
and we apply this value to all the output coupling efficiencies in our analysis.

\subsection{DC evaluation}
The measured insertion loss of our DCs is not negligible;
as such, instead of using the cross-bar ratio as described in Section~\ref{sec:DC evaluation} to obtain the DC cross coupling transmission,
we followed the regression method described in~\cite{zhang2024integrated} and obtained 
$T$\,=\,14.855\,±\,0.013\,dB. 

Nonetheless, we measured a cross-bar ratio of 0.03991\,$\pm$\,0.00035 
across our DCs, with a standard deviation of 0.9\%.
This is comparable to the repeatability of the measurements of a single waveguide to input and output lensed fibers (also 0.9\%),
and we thus conclude that our DCs are indeed uniform. 
We expect that the role of cross-bar ratio measurement in evaluating the DC uniformity would become more essential as the number of cascaded DCs scale to larger $n$.

\subsection{Responsivity evaluation}\label{sec:R_eval}

Using the pairwise method, we measured the responsivity $R$ of the integrated PDs using different numbers of DC stages $n$.
However, due to a light scattering noise floor of $\sim$\,50\,dB below the input fiber power $P$, our measurements are limited to a maximum of $n=3$ DC stages.
While this method does not strictly require the same input power for each measurement, we maintained an input fiber power of $P \sim 10$\,mW. 
The results are shown in table~\ref{tab for responsivity evaluation comparison} alongside that of a direct photocurrent measurement (Section~\ref{sec:direct}).

For the direct measurement, we observed that measurements at below $P$\,$\sim$\,0.4\,mW had increased uncertainties 
due to increased noise susceptibility at lower power levels.
The value reported here is measured at $P=0.4$\,mW, 
and we note that the measured responsivity does not deviate at higher power levels up to the few mW level. 

The measured responsivities from the pairwise and direct methods agree within the measurement uncertainties (see Section~\ref{sec:uncertainties} below).
However, the pairwise measurements significantly better repeatability (0.22\% error for $n=3$) compared to the direct photocurrent measurement (1.21\%). 
This does not worsen even after undergoing multiple stages of DC attenuation, 
which highlights the advantages of the pairwise method: 
besides the mitigation of input light power fluctuations by simultaneously measuring the optical power and photocurrent,
performing the optical measurements at higher power levels also results in higher signal-to-noise ratios and more repeatable fiber-to-chip coupling.

\begin{table}[htbp]
\centering
\caption{\bf PD Responsivity evaluation with pairwise measurements for varying stages of cascaded DCs $n$, and with a direct photocurrent measurement.}
\label{tab for responsivity evaluation comparison}
\begin{tabular}{|c|c|c|c|c|}
\hline
No. of cascaded DCs & $n$ = 1 & $n$ = 2 & $n$ = 3  & Direct       \\ \hline
PD responsivity (A/W) & 0.4844 & 0.4884 & 0.4928 & 0.5240\\ \hline
Repeatability error & 0.24\% & 0.13\% & 0.22\%  & 1.21\%      \\ \hline
\end{tabular}
\end{table}

\subsection{Uncertainty Analysis}\label{sec:uncertainties}
We performed an uncertainty analysis of the responsivity measurements, 
where the major uncertainty components include: 
measurements of the optical power, photocurrent, coupling losses, and the DC cross coupling, 
as well as the measurement repeatability. 

For the pairwise measurement method, according to the law of propagation of uncertainty \cite{iso1993guide}, 
we can derive the uncertainty of the responsivity $R$ from Eq.~(\ref{eq:for pair input reponsivity with cascaded DCs}):
\begin{align}
u_\textrm{pair}^2(R)&=\left[\frac{\partial R}{\partial {I_m}}\times u(I_m)\right]^2 + \left[\frac{\partial R}{\partial {I_n}}\times u(I_n)\right]^2 + \left[\frac{\partial R}{\partial {P'_{nm}}}\times u(P'_{nm})\right]^2\notag \\
&+ \left[\frac{\partial R}{\partial {P'_{mm}}}\times u(P'_{mm})\right]^2 + \left[\frac{\partial R}{\partial {C'_m}}\times u(C'_m)\right]^2 + \sum_{i=1}^{n-1} \left[\frac{\partial R}{\partial {T_i}}\times u(T_i)\right]^2
+ u^2(M)
\end{align}
where the partial derivative terms are the coefficients $c$ indicating the relative contributions of the respective uncertainty components, 
and $u(M)$ denotes the uncertainty contribution from the repeatability of the same measurement. 
Similarly, for the direct photocurrent measurement, we derive from Eq.~(\ref{eq:for single input responsivity}):
\begin{equation}
u_\textrm{direct}^2(R)=\left[\frac{\partial R}{\partial {I}}\times u(I)\right]^2 +
\left[\frac{\partial R}{\partial {P_D}}\times u(P_D)\right]^2 + \left[\frac{\partial R}{\partial {C'}}\times u(C')\right]^2
+ u^2(M)
\end{equation}

We provide below a detailed breakdown of the uncertainty components. 

A major uncertainty source in the optical power measurements is the optical power meter, which is calibrated with 1\% standard uncertainty.
The effects of the input optical power are canceled out for the pairwise method, but not for the direct photocurrent measurement.
We measured a power stability of 1\% for the 685\,nm laser, 
which is operated under current and temperature stabilization with a standard commercial mount and controller (Thorlabs).
The power measured from the input lensed fiber also has an added uncertainty of 1.7\%
due to the sensitivity of the measurement to even slight changes in the relative position and angle of the fiber end to the power meter.

The uncertainty of the photocurrent measurement is 0.05\%, given by the standard uncertainty of the sourcemeter instrument.
The uncertainty of the coupling efficiency terms $C'$ is 4.5\%,
dominated by the standard deviation in the measured fiber-to-chip coupling efficiencies.
The standard deviation of the DC cross coupling transmission $T$, as reported above, is 1.56\%.
The repeatability of the responsivity measurements are shown in Section~\ref{sec:R_eval}.

To obtain an expanded uncertainty, we consider the coverage factor $k$ which is applied as a numerical multiplier of the combined standard uncertainty; 
k\,=\,1 yields a standard uncertainty with level of confidence $p=68.27\%$, 
while we use k\,=\,2 for an expanded uncertainty of $p=95.45\%$ in our analysis.
Table~\ref{tab for uncertainty analysis} compares the uncertainty analysis of the PD responsivity obtained from with pairwise measurements with the largest attenuation ($n=3$ DC stages) and the direct photocurrent measurement.
For pairwise measurement method, with $n=1$, $n=2$ and $n=3$, 
the combined uncertainty of PD responsivity is 4.56\%, 4.82\% and 5.06\%, respectively, and the expanded uncertainty is 9.12\%, 9.63\% and 10.13\%, respectively.  
For the direct photocurrent measurement, the combined uncertainty is 5.16\%, and the expanded uncertainty is 10.32\%. 

\begin{table}[t!]
\centering
\caption{\bf Uncertainty analysis of the PD responsivity measurements with the pairwise and direct photocurrent methods. (u = uncertainty, c = coefficient)}
\label{tab for uncertainty analysis}
\begin{tabular}{|cc|cc|cc|}
\hline
\multicolumn{1}{|c|}{Uncertainty} & Affected 
& \multicolumn{2}{c|}{Pairwise, $n=3$}      & \multicolumn{2}{c|}{Direct}             \\ \cline{3-6}
\multicolumn{1}{|c|}{component} & quantity                                 & \multicolumn{1}{c|}{u}       & c   & \multicolumn{1}{c|}{u}          & c     \\ \hline
\multicolumn{1}{|c|}{\multirow{3}{*}{Optical power meter}}                 & $P'_{34}$    & \multicolumn{1}{c|}{1\%}     & 0.5 & \multicolumn{1}{c|}{-}          & -     \\
\multicolumn{1}{|c|}{}                                              & $P'_{44}$    & \multicolumn{1}{c|}{1\%}     & 0.5 & \multicolumn{1}{c|}{-}          & -     \\
\multicolumn{1}{|c|}{}                                              & $P_D$     & \multicolumn{1}{c|}{-}       & -   & \multicolumn{1}{c|}{1\%}        & 1     \\ \hline
\multicolumn{1}{|c|}{Input optical power}                                   & $P_D$     & \multicolumn{1}{c|}{-}       & -   & \multicolumn{1}{c|}{1.7\%}     & 1     \\ \hline
\multicolumn{1}{|c|}{\multirow{3}{*}{Sourcemeter}}                 & $I_3$     & \multicolumn{1}{c|}{0.05\%}  & 0.5 & \multicolumn{1}{c|}{-}          & -     \\
\multicolumn{1}{|c|}{}                                              & $I_4$     & \multicolumn{1}{c|}{0.05\%}  & 0.5 & \multicolumn{1}{c|}{-}          & -     \\
\multicolumn{1}{|c|}{}                                              & $I$      & \multicolumn{1}{c|}{-}       & -   & \multicolumn{1}{c|}{0.05\%}     & 1   \\ \hline
\multicolumn{1}{|c|}{Coupling efficiency}                           & $C'_4, C'$ & \multicolumn{1}{c|}{4.5\%}   & 1   & \multicolumn{1}{c|}{4.5\%}      & 1     \\ \hline
\multicolumn{1}{|c|}{\multirow{2}{*}{DC cross transmission factor}} & $T_1$     & \multicolumn{1}{c|}{1.56\%}  & 1   & \multicolumn{2}{c|}{\multirow{2}{*}{-}} \\
\multicolumn{1}{|c|}{}                                              & $T_2$     & \multicolumn{1}{c|}{1.56\%}  & 1   & \multicolumn{2}{c|}{}                   \\ \hline
\multicolumn{1}{|c|}{Measurement repeatability}                     & $M$      & \multicolumn{1}{c|}{0.22\%}  & 1   & \multicolumn{1}{c|}{1.21\%}     & 1     \\ \hline
\multicolumn{2}{|r|}{Combined uncertainty}                                                       & \multicolumn{1}{c|}{5.06\%}  &     & \multicolumn{1}{c|}{5.16\%}     &       \\
\multicolumn{2}{|r|}{Expanded uncertainty (k\,=\,2)}                                               & \multicolumn{1}{c|}{10.13\%} &     & \multicolumn{1}{c|}{10.32\%}    &       \\ \hline
\end{tabular}
\end{table}

\section{Discussion}
At first glance, the expanded uncertainty for the evaluation of the PD responsivity $R$ using the pairwise measurement method (10.13\%) 
does not appear to be a significant improvement over the direct photocurrent measurement (10.32\%).
However, we argue that these results do not reflect the full potential of the pairwise method with cascaded DCs.
We will discuss below the advantages of the pairwise method, 
as well as how the major uncertainty components can be improved towards single-photon-level optical power levels. 

\subsection{Advantages of the pairwise method with cascaded DCs}
There are several systematic advantages of the pairwise measurement method with cascaded DCs over a direct photocurrent measurement.
We have already demonstrated better measurement repeatability with the pairwise method (0.22\% vs 1.21\%).
Going further towards lower optical power levels, 
we expect that the direct measurement of very weak input optical power levels will have worse signal-to-noise ratio and hence even larger uncertainties for the direct method.
In contrast, the pairwise method measures optical powers that do not undergo the full attenuation, 
and thus are still at relatively high power levels (near $\sim$\,mW) and can be measured with lower errors.

The simultaneous measurement of the PD photocurrent and the optical power essentially functions as a built-in power monitor, 
thus mitigating the effect of power drifts or fluctuations. 

Despite the pairwise method involving more measurements with multiple waveguides, the input coupling terms $C_n$, $C_m$ are systematically canceled out and thus do not incur additional uncertainties.

Lastly, the DCs allow for the in-situ measurement of the coupling efficiencies directly without additional photonics structures, 
which is not possible with only a single direct input waveguide.

\subsection{Reducing uncertainties}
From the uncertainty analysis presented in Table~\ref{tab for uncertainty analysis}, 
a major uncertainty component of the pairwise measurements 
was the DC cross transmission factor $T$, with an uncertainty of 1.56\%.
This is largely limited by the scattering noise floor, which limited the number of cascaded DC stages we could include in our regression analysis to $n=3$. 
This can be significantly improved in future device iterations 
with light-shielding structures (metallic or otherwise) surrounding the PD 
that would block and mitigate the effect of any scattered light on-chip. 
We expect that an improvement in the uncertainty of $T$ to match that of the the measurement of the cross-bar ratio ($\sim$\,0.9\%) is practically achievable.

Our devices use inverse taper edge coupler structures with a minimum taper design width of 250\,nm, 
and their small feature size makes them susceptible to
non-uniformity across devices.
We thus expect that the coupling efficiency uncertainty of 4.5\%, 
which is the largest component in our uncertainty analysis,
can be improved to $<$1\% with further process optimization.

For the target application of on-chip single-photon detector calibration,
we consider how we can obtain single-photon levels ($\sim$\,aW) from $\sim$\,mW light sources.
Assuming the same DC design presented here with a nominal attenuation of $\sim$\,15\,dB per stage, 
we can extend our attenuation circuit to $n=10$ cascaded DC stages with a total attenuation of $\sim$\,150\,dB.
With the projected improvements in the uncertainty components as described above,
the pairwise method would yield an expanded uncertainty of $<$6\%.

\section{Conclusion}

We have demonstrated a photonics integrated attenuation circuit 
based on cascaded directional couplers (DCs)
for evaluating the responsivity of on-chip photodetectors (PDs).
With simultaneous pairwise measurements of the integrated PD photocurrent and an auxiliary optical output, 
we obtained a measurement repeatability error of 0.22\% and an overall expanded uncertainty (k\,=\,2) of 10.13\%
using 3 cascaded DC stages.
We have discussed the advantages of the pairwise method, 
including the systematic mitigation of various optical power uncertainty components,
as well as how the other uncertainty contributions can be improved.
Our scheme holds great potential to be extended to the characterization of integrated single-photon detectors with an on-chip photonics calibration circuit.

\begin{backmatter}
\bmsection{Funding}
We acknowledge the funding from the Agency for Science, Technology and Research, Singapore (C230917005),
and the National Research Foundation, Singapore, under the Quantum Engineering Programme (W21Qpd0304).

\bmsection{Disclosures}
The authors declare no conflicts of interest.

\bmsection{Data Availability Statement}
Data underlying the results presented in this paper may be obtained from the authors upon
reasonable request.
\end{backmatter}

\clearpage
\bibliography{main}

\end{document}